\documentclass[
    ,final            % use final for the camera ready runs
  ]
  {aipproc}
\layoutstyle{6x9}
\usepackage{graphicx}
\input mysym
\begin{document}
% remove the following for publication
\begin{figure}
%
%\leftline{\includegraphics[scale=0.5]{cdfii_thumb_logo.eps}\hfill
%CDF/PHYS/BOTTOM/PUBLIC/7347 }
\leftline{\hfill FERMILAB-CONF-06-002-E}
\end{figure}
\title{Heavy Flavor Production in CDF II Detector}
\classification{13.87.Ce, 13.85.Ni, 13.85.Qk, 13.87.-a, 13.87.Fh}
\keywords      {Heavy Quark, Jet, Production}
\author{Igor V. Gorelov\thanks{talk given on behalf of the CDF
                               Collaboration at the XVII Particles 
                               and Nuclei International Conference,
                               PANIC~05,~October~24-28,~2005,
                               ~Santa~Fe,~New~Mexico.}\\
       (For the CDF Collaboration)}
       {address={Department of Physics and Astronomy,\\ 
         University of New Mexico,\\ 
         800 Yale Blvd. NE, Albuquerque, NM 87131, USA\\
         email:gorelov$@$fnal.gov}
}
\begin{abstract}
  For data collected with the CDF Run II detector,
  measurements of the charm and bottom production cross-sections are
  presented. The results are based both on large samples of fully
  reconstructed hadron decay products of charm and bottom made
  available by the tracking triggers and on a calorimeter jet triggered
  sample tagged by the presence of a secondary vertex. The experimental 
  data are compared with theoretical predictions from recent 
  next-to-leading order (NLO) QCD calculations.
\end{abstract}
\maketitle
%
%%%%%%%%%%%%%%%%%%%%%%%%%%%%%%%%%%%%%%%%%%%%
%% MAINMATTER
%%%%%%%%%%%%%%%%%%%%%%%%%%%%%%%%%%%%%%%%%%%%
%
%\section{Status}
%
 \par	
 The CDF Run I measurements~\cite{CDF:Run1} of the \b-quark production cross-section
 compared with next-to-leading order (NLO) QCD theoretical predictions
 \cite{th:Nason89,th:Mangano97} yielded a ratio 
 \( Data/Theory = 2.9\pm0.2({stat\oplus\,syst_{\pt}})\pm0.4({syst_{fc}})\)
(see~\cite{CDF:Run1} for details). 
 The apparent excess in the data over theoretical expectations had been
 established. The controversy has been subsequently re-examined by 
 experimental and theoretical parties. The upgraded CDF and D0
 detectors at the Tevatron entered the \run2 data-taking period. A
 dramatic theoretical advance also occurred: a new
 theoretical technique was developed 
 \cite{th:Mangano97,th:Cacciari98,th:Cacciari02}, which implemented a
 re-summation at the next-to-leading-log (NLL) level. The formalism was
 applied both to a fixed order perturbative calculation and to
 extraction of the non-perturbative part of the fragmentation function from
 previous experimental LEP data~\cite{ALEPH:bfrag,ALEPH:dstfrag}. As a result a
 $Data/Theory$ ratio (still for the CDF Run I data) 
 was decreased from  $2.9$ to
 \(1.7\pm0.5(exp)\pm0.5(theory)\)\cite{CDF:Run1,th:Cacciari02}.
%
%\section{\D-mesons}
%
 \par
\begin{figure}[!ht]
 \resizebox{.9\textwidth}{!}
  {\includegraphics[draft=false]{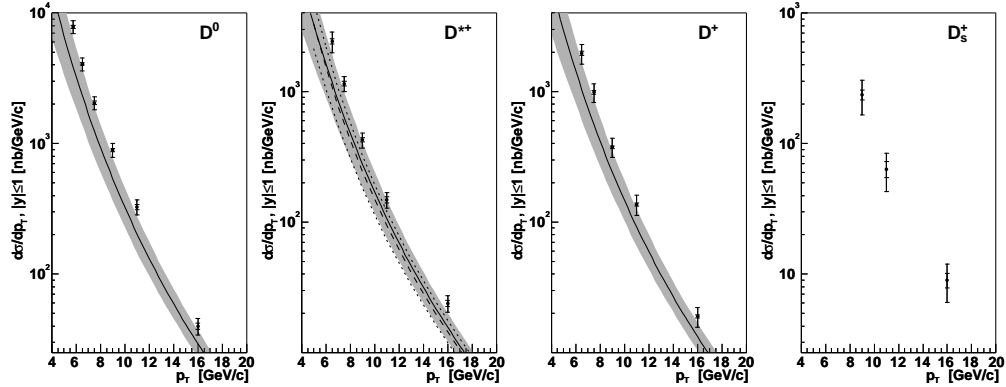}}
 \caption{ The differential \Dz,\,\Dp,\,\Dstarp and \Ds cross-section 
           distributions measured~\cite{CDF:cprod} with the \cdf2 detector. 
           The theoretical predictions are shown as a shaded band. }
 \label{fig:dmesons}
\end{figure}

 
 The upgraded tracking of the \cdf2 detector provided an excellent
 opportunity for heavy-quark physics measurements. The fine track impact parameter
 ($d_0$) resolution of the CDF silicon tracker SVX-II made it possible to
 implement a new displaced track trigger selecting events
 with a pair of high \pt and large $d_0$ tracks. This trigger yielded
 a large data sample dominated by charm states decaying through hadron
 modes including \DzKpi, \DpKpipi, and \DstarDpi \footnote{Unless otherwise
 stated all references to the specific charge combination imply the
 charge conjugate combination as well.}. CDF has
 measured~\cite{CDF:cprod} the direct production cross-sections of
 \D-meson species. As both prompt and secondary (from \b-quark decays)
 \D-mesons contribute to the visible yield, the prompt fraction of the
 signals was extracted by fitting the shape of the impact parameter
 distribution for each particular hadron mode. The experimental
 results shown at Figure~\ref{fig:dmesons} have been compared with
 fixed order NLL (FONLL) calculations made for the charm sector~\cite{th:Cacciari03}. 
 The agreement seems to be moderate, with data lying at the upper bound of
 the theoretical uncertainties.
%
%
%\section{\b-Hadrons}
%
 \par
\begin{figure}[!ht]
 \resizebox{.5\textwidth}{!}
  {\includegraphics[draft=false]{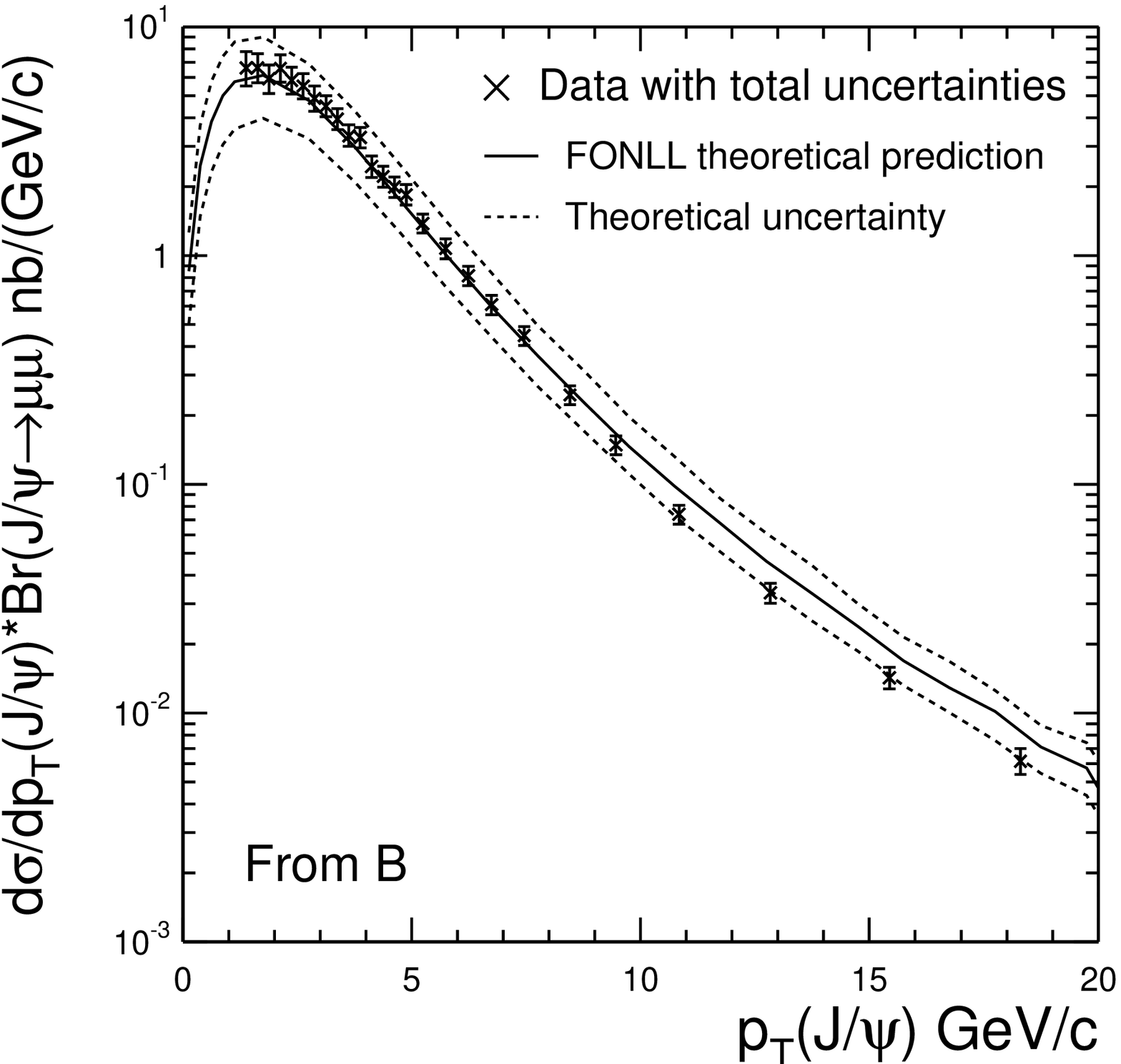}}
 \caption{ The differential cross-section for 
           \( \pap\to{H_b\X},H_b\to\jpsi\X,\,\textrm{and}\,\,\jpsi\to\mumu \) 
           with |y(\jpsi)|<0.6 and \pt(\jpsi)>1.25\gevc.
           The theoretical predictions are outlined in~\cite{th:Cacciari04}. 
           The central theoretical values underestimate the data points 
           below $\sim7.5\gevc$ and overestimate the ones above $\sim7.5\gevc$.
 }
 \label{fig:bjpsi}
\end{figure}

	
 Another tracker-based trigger exploits the CDF central tracking chamber,
 for which reconstructed tracks are matched with hits in the muon chambers.
 At the highest Level-3 trigger a pair of muon candidates having
 opposite charge is reconstructed, and only pairs with an invariant
 mass \( M(\mumu)\in[2.7, 4.0]\gevcc \) around the mass of the \jpsi are
 accepted. This large triggered data sample of \jpsi mesons is used to
 measure the inclusive cross-section for \b-hadrons, $H_b\to\jpsi X$ . The
 \jpsi from the decay of $H_b$ is likely to be displaced from the
 primary vertex where \b-hadrons are assumed to be produced. The
 maximum likelihood fit of the pseudo-proper time \( c\tau =\lxy\cdot(M/\pt) \) 
 in bins of \pt(\jpsi) finds the \b-fraction $f_{b}$ of \jpsi yield in every
 \pt bin. The measured~\cite{CDF:jpsiprod} differential \b-hadron cross-section spectrum
 is shown in Figure~\ref{fig:bjpsi} where it is compared with FONLL
 calculations~\cite{th:Cacciari04}. The integrated total cross-section 
 was found to be 
 \( \sigma(\pap\to\,H_b,H_b\to\,\jpsi,\pt(\jpsi)>1.25\gevc,|y(\jpsi)|<0.6) 
 = 0.330\pm0.005(stat)^{+0.036}_{-0.033}(syst)\,\mub \).
%
%\section{\b-Jets}
%
 \par
\begin{figure}[!ht]
 %\resizebox{.6\textwidth}{!}
 % {\includegraphics[draft=false]{figs/bcross_nlo_comp_escale.eps}}
 \resizebox{.8\textwidth}{!}
  {\includegraphics[draft=false]{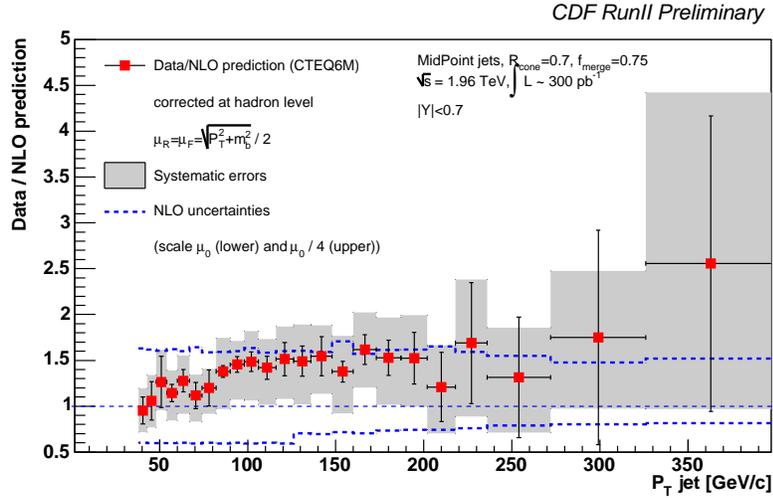}}
 \caption{The comparison of \b-jet cross-section measurements 
          with NLO predictions by S. Frixione~\etal made at the hadron level.
         }
 \label{fig:bjet}
\end{figure}

 
 Heavy quark jets are good observables as their experimental
 energy spectra do not depend on knowledge of heavy-quark
 fragmentation functions, contrary to the cases discussed above.  The
 systematics due to decay properties of heavy hadrons are reduced as
 well. The predictions of NLO models are expected to be more precise
 and less sensitive as collinear gluons causing large $\log$-s are
 counted in jet cones.  The \cdf2 \b-jet cross section measurements
 are made with a $\sim300\invpb$ data sample triggered by inclusive jet
 triggers based on the \cdf2 calorimeters. The \b-jet production study
 extends the \pt reach beyond the [1.25, 25]\gevc range accessible with final
 heavy hadron modes.  The central jets of rapidity range
 $|Y_{jet}|<0.7$ are reconstructed in Y-$\phi$ space using a
 cone-based iterative MidPoint algorithm~\cite{th:midpoint} 
 with a cone radius $R_{cone}=0.7$.
 The hadronic energy scale of the jets is corrected and unfolded to the 
 hadron particle
 level using Monte Carlo (MC) data samples. Additional correction is applied 
 for ambient energy flow into the jet cone due to the underlying event.
 As \b-quark decay products within the jet cone are also likely to be
 displaced from the primary vertex, reconstructed tracks are matched
 with the jet cone and required to have a displaced impact parameter
 $d_0$. Furthermore these tracks are fitted to a common secondary
 vertex, and the corresponding secondary vertex decay length is
 required to be significantly above zero, 
 \( \lxy/\sigma_{\lxy}>3.0 \). 
 The jet satisfying these conditions is positively tagged as \b-jet candidate. 
 The tagged jet is likely to contain a decay
 vertex of \b-, \c- or light quarks (e.g. \s- quark).  Hence again the
 \b-fraction $f_{b}$ must be extracted. A fit of the reconstructed 
 secondary vertex invariant mass
 distribution to MC templates of \b- and
 non-\b components of positively tagged jets finds an $f_{b}$ for
 tagged jets in every \pt bin.
 The comparison of \cdf2 preliminary \b-jet cross-section
 measurements to NLO predictions by S. Frixione~\etal (see
 e.g.~\cite{th:Frixione96}) with parton distribution function (PDF)
 of version CTEQ6M is shown in Figure~\ref{fig:bjet}. 
 The statistical and systematic errors on the last bins are dominated by the
 error on the b-tagged jets fraction. One of the main systematic error
 contributions comes from the jet energy scale.
 \par 
 In conclusion, \cdf2 has measured inclusive production cross-sections in the central
 rapidity region at $\sqrt{s}=1960\gev$ for \D-mesons in
 exclusive decay modes with $\pt(\D)>5.5\gevc$, for \b-hadrons in decay
 modes containing \jpsi in a \pt range of $\sim0$ to 20\gevc. The
 data and some recent theoretical calculations show reasonable
 agreement within experimental and theoretical uncertainties. 
 \cdf2 has presented preliminary results on central
 production of \b-tagged jets in a wide range, from 38 to
 360\gevc.  The measurements are also in agreement with the most
 recent NLO predictions given theoretical and experimental
 uncertainties.
%
%%%%%%%%%%%%%%%%%%%%%%%%%%%%%%%%%%%%%%%%%%%%%%%%
%% BACKMATTER
%%%%%%%%%%%%%%%%%%%%%%%%%%%%%%%%%%%%%%%%%%%%%%%%
%
\begin{theacknowledgments}
  The author is grateful to his colleagues from the CDF $B$-Physics
  and QCD Working Groups for useful suggestions
  and comments made during preparation of this talk. The author would
  like to thank Prof.~Sally~C.~Seidel for support of this work,
  fruitful discussions, and comments. 
\end{theacknowledgments}

\bibliographystyle{aipproc}   % if natbib is available
%% The following lines show an example how to produce a bibliography
%% without the help of the BibTeX program. This could be used instead
%% of the above.
%%%%%%%%%%%%%%%%%%%%%%%%%%%%%%%%%%%%%%%%%%%
%

%
%%%%%%%%
\end{document}